\documentstyle[12pt]{article}

\begin{document}

\title{Dynamical symmetry breaking and the Nambu-Goldstone theorem \\
in the Gaussian wave functional approximation}
\author{V. Dmitra\v sinovi\' c $^{1}$ and Issei Nakamura $^{2}$ \\
$^{1}$ Vin\v ca Institute of Nuclear Sciences \\
P.O.Box 522, 11001 Beograd, Yugoslavia \\
$^{2}$ Research Center for Nuclear Physics, Osaka University \\
Ibaraki, Osaka 567-0047, Japan }

\maketitle
\abstract{We analyze the group-theoretical ramifications of the
Nambu-Goldstone [NG] theorem in the
self-consistent relativistic variational Gaussian wave functional
approximation to spinless field theories. In an illustrative example
we show how the Nambu-Goldstone theorem would work in the O(N)
symmetric $\phi^4$ scalar field theory,
if the residual symmetry of the vacuum were lesser than O(N-1), e.g.
if the vacuum were O(N-2), or O(N-3),... symmetric.
[This does not imply that any of the ``lesser" vacua is actually the
absolute energy minimum: stability analysis has not been done.]
The requisite number of NG bosons would be (2N - 3), or (3N - 6), ...
respectively,
which may exceed N, the number of elementary fields in the Lagrangian.
We show how the requisite new NG bosons would appear even in channels
that do not carry the same quantum numbers as one of N ``elementary
particles" (scalar field quanta, or
Castillejo-Dalitz-Dyson [CDD] poles) in the Lagrangian,
i.e. in those ``flavour" channels that have no CDD poles.
The corresponding Nambu-Goldstone bosons are composites (bound states)
of pairs of massive elementary (CDD) scalar fields excitations.
As a nontrivial example of this method we apply it to the
physically more interesting
't Hooft $\sigma$ model (an extended $N_{f} = 2$ bosonic linear
$\sigma$ model with four scalar and four pseudoscalar fields), with
spontaneously
and explicitly broken chiral $O(4) \times O(2) \simeq SU_{\rm R}
(2) \times SU_{\rm L}(2) \times U_{\rm A}(1)$ symmetry.}

\section{Introduction}
\label{s:intro}

The proof of the Nambu-Goldstone [NG] theorem
\cite{njl61,gold61,gsw62,klein63,gilbert64} in the
Gaussian wave functional approximation
\cite{schiff63,rosen68,cjt74,bg80,sym81,stev84,stev87,hat92}
used to be an open problem for over 30 years,
see Refs. \cite{ku64,hung66,bc85,dms96}.
The first proof was given in the O(2)
symmetric $\phi^4$ theory \cite{dms96}, and then straightforwardly
extended to O(4) in \cite{issei01}; the crucial assumption was that
the ground state be O(N-1) symmetric, i.e. that only one (scalar)
field develops a vacuum expectation value [v.e.v.]. \footnote{That
assumption
is justified when the remaining N-1 fields are pseudoscalars, as in
the Gell-Mann--Levy model, otherwise CP symmetry is (spontaneously)
broken.}
By the standard NG boson counting methods \cite{gsw62,klein63},
for every spontaneously broken symmetry Lie group generator there is
one NG boson.
As the Lie algebra O(N) has $N(N-1)/2$ generators, for a ground state
(vacuum) with an O(N-1) residual symmetry the number of NG bosons ought
to be $N(N-1)/2 - (N-1)(N-2)/2 = N-1$. That is exactly the number of
available fields in the Lagrangian. What happens when the residual
symmetry the ground state is ``smaller" than O(N-1) and there ought
to be more than N-1 NG bosons?

In this paper we shall extend our proof of the NG theorem in the
Gaussian approximation \cite{dms96,issei01}to the O(N) symmetric
$\phi^4$ theory
when the symmetry of the ground state is dynamically broken to some
(proper) subgroup of O(N-1), in this specific case to one of the
following symmetries: $O(N-2) \times O(2)$,
$O(N-3) \times O(3)$, ... , $O(N/2) \times O(N/2)$ for N even,
or $O((N-1)/2) \times O((N+1)/2)$ for N odd.
As we shall show, the residual symmetry pattern is dictated by
the absolute minimum of the O(N) $\phi^4$ model's ground state energy,
which in turn depends on the dynamics (gap equation) in the Gaussian
approximation, as well as the free parameters. Absolute minimization
of the vacuum energy will not been done in this paper, only a
search for the extremal and/or saddle points.
In a different model or approximation, or for different values of
the free parameters the residual symmetry might be different.
In any case, with such an asymmetric ground state the number of
NG bosons must exceed N-1, the largest number of ``elementary"
scalar fields available in the Lagrangian
(at least one field must develop the vacuum expectation value
and thus cannot create or destroy single NG bosons).
Nevertheless the canonical number of massless spinless excitations
appears in the spectrum \cite{klein63}.
Our proof ought to leave no doubt as to the
composite nature of the NG bosons in the Gaussian approximation.

This paper falls into five Sections and two Appendices.
Firstly in Sect. \ref{s:GA} we define the O(N) $\phi^{4}$ model and
the Gaussian approximation. In Sect. \ref{s:NG} we show
exactly how the requisite number of massless NG (bosonic bound) states
appear and that all the ``broken symmetry" N\" other currents
remain conserved, as the symmetry of the vacuum is reduced.
NG bosons invariably appear in the Gaussian approximation to the
O(N) $\phi^{4}$ model in those channels that
also contain CDD poles, so
in Sect. \ref{s:th} we extend our proof to the
't Hooft $\phi^{4}$ model which does not have this property.
There we show how various bound NG states appear, or disappear as the
symmetries of the ground state and/or the Lagrangian change.
Finally, in Sect. \ref{s:s&c} we draw our conclusions and set them
in a wider context.
In the Appendices we present some technical details omitted in the
main body of text.

\section{Preliminaries}
\label{s:GA}

\subsection{The O(N) symmetric scalar $\phi^4$ model}
\label{s:O(N)}

At first we confine ourselves to the $O(N)$ symmetric
scalar $\phi^4$ theory for the sake of simplicity.
All scalar field theories with other spontaneously
broken internal symmetries can be reduced to some
subgroup of O(N). Of course, in such cases there
will be interaction terms other than the simple $\phi^4$
one shown below.
The Lagrangian density of this theory is
\begin{eqnarray}
{\cal L} &=&  \frac{1}{2} \left(\partial_{\mu}
\mbox{\boldmath$\phi$}\right)^{2} -
V(\mbox{\boldmath$\phi$}^{2}) ~,\
\label{e:lag}
\end{eqnarray}
where
$$\mbox{\boldmath$\phi$} =
(\phi_{0},\phi_{1},\phi_{2}, ... , \phi_{N-1})
= (\sigma, \mbox{\boldmath$\pi$}),$$
is a column vector and
$$V(\mbox{\boldmath$\phi$}^{2}) = - {1 \over 2} \mu_{0}^{2}
\mbox{\boldmath$\phi$}^{2} + {\lambda_{0} \over 4}
\left(\mbox{\boldmath$\phi$}^{2}\right)^{2}. $$
We assume here that $\lambda_{0}$ and $\mu_{0}^{2}$ are not only
positive, but such
that spontaneous symmetry breakdown (SSB) occurs in the Gaussian
approximation [GA] introduced below.

\subsection{The Gaussian variational method}
\label{s:gauss}

\subsubsection{The Gaussian ground state (``vacuum")}
\label{s:gaen}

The Rayleigh-Ritz variational approximation to quantum field theories
is based on the (``elliptical") Gaussian ground state (vacuum)
functional Ansatz, Refs. \cite{bg80,sym81,stev84,hat92},
\begin{eqnarray}
\Psi_{0}[{\vec \phi}] = {\cal N} \exp \left( - {1 \over 4 \hbar} \int
d{\bf x} \int d{\bf y} \left[\phi_{i}({\bf x}) - \langle  \phi_{i}
({\bf x}) \rangle \right]
G_{ij}^{-1}({\bf x},{\bf y})\left[\phi_{j}({\bf y}) - \langle \phi_{j}
({\bf y})) \rangle \right]\right)~,
\label{e:gaus}
\end{eqnarray}
where ${\cal N}$ is the normalization constant, and one sums
all repeated (roman lettered) indices from 0 to N-1.
$\langle \phi_{i}({\bf x})\rangle$
is the vacuum expectation value (v.e.v.) of the $i$-th spinless field
which henceforth we will assume to be translationally invariant
$\langle  \phi_{i}({\bf x}) \rangle =
\langle  \phi_{i}(0) \rangle \equiv \langle  \phi_{i} \rangle$
and
$$G_{ij}({\bf x},{\bf y}) =
{1 \over 2} \delta_{ij} \intop\limits {d {\bf k}
\over (2 \pi)^{3}} {1 \over \sqrt{{\bf k}^{2} + m_{i}^{2}}}
e^{i {\bf k} \cdot ({\bf x} - {\bf y})} .$$
We have explicitly kept $\hbar$ (while setting
the velocity of light $c = 1$) to keep track of quantum corrections
and count the number of ``loops" in our calculation.
Then the ``vacuum'' (ground state) energy density becomes
\begin{eqnarray}
{\cal E}(m_{i}, \langle  \phi_{i} \rangle) &=&
- {1 \over 2} \mu_{0}^{2} \langle \mbox{\boldmath$\phi$} \rangle^{2}
+ {\lambda_{0} \over 4}
\left[\langle \mbox{\boldmath$\phi$} \rangle^{2}\right]^{2}
\nonumber \\
&+&
\hbar \sum_{i =0}^{N-1} \left[I_{1}(m_{i}) - {1 \over 2}
(\mu_{0}^{2} +
m_{i}^{2}) I_{0}(m_{i}) \right]
\nonumber \\
&+&
{\lambda_{0} \over 4} \Bigg\{
6 \hbar \sum_{i=0}^{N-1} \langle \phi_{i} \rangle^{2} I_{0}(m_{i}) +
2 \hbar \sum_{i \neq j = 0}^{N-1} \langle \phi_{i} \rangle^{2}
I_{0}(m_{j})
\nonumber \\
&+&
\hbar^{2} \sum_{i=0}^{N-1} I_{0}^{2}(m_{i})
+ \hbar^{2} \sum_{i \neq j = 0}^{N-1} I_{0}(m_{i}) I_{0}(m_{j}) \Bigg\} \ ~,
\label{e:ener}
\end{eqnarray}
where
\begin{eqnarray}
I_{0}(m_{i}) &=& {1 \over 2} \int {d{\bf k} \over (2 \pi)^{3}}
{1 \over \sqrt{{\bf k}^{2} + m_{i}^{2}}} =
i \int {d^{4} k \over (2 \pi)^{4}}
{1 \over {k^{2} - m_{i}^{2} + i \epsilon}} =
G_{ii}({\bf x},{\bf x})
\label{e:I_0} \\
 I_{1}(m_{i}) &=& {1 \over 2} \int {d {\bf k} \over (2
\pi)^{3}}
\sqrt{{\bf k}^{2} + m_{i}^{2}} =
- {i \over 2} \int {d^{4} k \over (2\pi)^{4}}
\log \left(k^{2} - m_{i}^{2} + i \epsilon \right) +
{\rm const.}~.
\label{e:I_1} \
\end{eqnarray}
We may identify $\hbar I_{1}(m_{i})$ with the familiar ``zero-point"
energy density of a free scalar field of mass $m_{i}$.

The divergent integrals $I_{0,1}(m_{i})$ are understood to be regularized
via an UV momentum cutoff $\Lambda$. Thus we have introduced a new
free parameter into the calculation. This was bound to happen in one
form or another, since even in the renormalized perturbation theory one must
introduce a new dimensional quantity (the ``renormalization scale/point'')
at the one loop level. We treat this model as an effective theory and
thus keep the cutoff without renormalization \footnote{There are
several renormalization schemes for the Gaussian approximation, but
they show signs of instability and ultimately
seem to lead to ``triviality" \cite{stev87}.}.

\subsubsection{The vacuum energy minimization equations}
\label{sec:gap1}

We vary the energy density with respect to the field vacuum expectation
values $\langle \phi_{i} \rangle$ and the ``dressed" masses $m_{i}$.
The extremization condition with respect to the field vacuum
expectation values (v.e.v.) reads:
\begin{eqnarray}
\left({\partial {\cal E}(m_{i}, \langle  \phi_{i} \rangle) \over
\partial
\langle \phi_{j} \rangle}\right)_{min} = 0; ~~~j \in
0,1,\ldots, N-1~;
\label{e:min1}
\end{eqnarray}
or explicitly
\begin{eqnarray}
\langle  \phi_{j} \rangle \left[- \mu_{0}^{2} + {\lambda_{0}}
\left(\langle \mbox{\boldmath$\phi$} \rangle^{2}
+ 3 \hbar I_{0}(m_{j})
+ \hbar \sum_{j \neq i=0}^{N-1} I_{0}(m_{i}) \right) \right]_{min}
&=& 0~. \
\label{e:vac}
\end{eqnarray}
The second set of energy extremization equations reads
\begin{eqnarray}
\left({\partial {\cal E}(m_{i}, \langle  \phi_{i} \rangle)
\over \partial m_{j}}\right)_{min} = 0, ~~~j \in 0,1, \ldots ,N-1~,
\label{e:min2}
\end{eqnarray}
or
\begin{eqnarray}
m_{j}^{2} + \mu_{0}^{2} &=&
{\lambda_{0}} \left(3 \langle \phi_{j}^{2} \rangle +
\hbar \sum_{j \neq k=0}^{N-1} \langle \phi_{k}^{2} \rangle
+ 3 \hbar I_{0}(m_{j}) +
\hbar \sum_{j \neq k=0}^{N-1} \hbar I_{0}(m_{k}) \right)_{min}
.\
\label{e:1-b}
\end{eqnarray}
Eqs. (\ref{e:vac}),(\ref{e:1-b}) can be identified with the
(truncated) Schwinger-Dyson (SD) equation \cite{hat92,chai79}
for the one- and two-point Green functions, see Ref.
\cite{dms96,issei01}.
The solutions to these equations plus the additional
minimization requirements (positive-definite second derivatives
matrix, i.e., the positive definiteness of its principal minors,
which will not be discussed here)
determine the symmetry of the ground state (vacuum).

\subsubsection{The vacuum symmetry}
\label{s:vacsym}

Now one ordinarily assumes \footnote{The validity of this assumption,
of course, depends on the values of the bare parameters in the
Lagrangian and the cutoff $\Lambda$, or the values of
the renormalized parameters, if one insists on renormalization.
One may also have an unbroken symmetry: with all the fields having a
zero v.e.v.,
$\langle \phi_{i} \rangle = 0, i \in 0,1, ...,N-1$, the gap
equations (\ref{e:vac}),(\ref{e:1-b}) lead to all the masses $m_{i}$
being equal. In other words the N fields form an O(N) multiplet, so we
may say that the symmetry of the vacuum is O(N), i.e. not broken.}
that only one of the scalar fields $\phi_{i}$
[by convention the $i = 0$ one] develops a nonzero v.e.v., i.e.
$\langle \phi_{0} \rangle \neq 0$ and then one proceeds with the
proof of the NG theorem.

In the conventional case the scalar field masses are
$m_{0} = M; m_{1} = m_{2} = m_{3} = ... = \mu$, i.e.,
the (N - 1) fields $\phi_{i}; i \in 1,2,..., N-1$, of mass
$m_{i} = \mu$, form an (N - 1)-plet
and the residual symmetry of the vacuum is O(N-1).
That, however, is not the only logical possibility: one may
assume that more than one field develops v.e.v..
This, of course, means that the (residual)
symmetry of the vacuum is ``lesser" than the one
in the (``canonical") case with only one v.e.v. \footnote{
In the Born approximation such a vacuum might be reducible to
the ``one v.e.v." vacuum by means of an O(N) transformation
(if it lies in the same ``orbit" of the symmetry group
\cite{cicogna}), but in the Gaussian approximation dynamical symmetry
breaking may lead to an irreducibly different ground state.}.

The decision which of these possibilities actually takes place
can be made on the basis of comparing their respective ground state
energies. In the case with an O(N-1) symmetric ground state, the
vacuum energy, or effective potential is
\begin{eqnarray}
{\cal E}(M, \mu, \langle  \phi_{0} \rangle)_{min} &=&
- {1 \over 2} \mu_{0}^{2} \langle \phi_{0}
\rangle^{2}
+ {\lambda_{0} \over 4}
\langle \phi_{0}
\rangle^{4}
\nonumber \\
&+&
\hbar
\left[I_{1}(M) - {1 \over 2}(\mu_{0}^{2} + M^{2}) I_{0}(M) \right]
\nonumber \\
&+&
(N-1)\hbar
\left[I_{1}(\mu) -
{1 \over 2}(\mu_{0}^{2} + \mu^{2}) I_{0}(\mu)\right]
\nonumber \\
&+&
{\lambda_{0} \over 4} \Bigg\{
2 \hbar \langle \phi_{0} \rangle^{2} \left[3 I_{0}(M) +
(N-1) I_{0}(\mu) \right] + 3 \hbar^{2} I_{0}^{2}(M)
\nonumber \\
&+&
(N-1) \hbar^{2} I_{0}(\mu) \left[2 I_{0}(M) + (N+1) I_{0}(\mu)\right]
\Bigg\} \ ~.
\label{e:ener1}
\end{eqnarray}
Similar expressions for ground states with lesser symmetries than
O(N-1) can be derived by applying the corresponding gap equations
(\ref{e:vac}),(\ref{e:1-b}) to Eq. (\ref{e:ener}). The question of
the {\it absolute} energy minimum will not be pursued in this paper,
however.
For the purpose of argument we shall assume that alternative ground
states
exist and are stable, i.e., energetically favourable to the standard
one.

If two (or more) scalar  fields' v.e.v. are simultaneously nonzero,
e.g. if
$\langle \phi_{0} \rangle = v_{1} \neq 0$ and
$\langle \phi_{j} \rangle = v_{2} \neq 0; j \in 1,2,...,N-1$
$\langle \phi_{i} \rangle = 0; i \in 1,2, ...,j-1,j-2,...,N-1$,
we may change the scalar field labels such that
the fields with nonzero v.e.v.s are labeled successively  0,1,...,k
without loss of generality.
In such a case the masses $m_{i}$ may also take on different values.

For simplicity's sake we assume that only two fields
have nonvanishing vacuum expectation values (v.e.v.)
\footnote{Extension
to three, or more v.e.v.s follows by straightforward analogy.},
$\phi_{0}, \phi_{1} \neq 0$ (we label their masses
$m_{0} = M, m_{1} = \mu_{1}$).
The remaining (N - 2) fields $\phi_{i}; i \in 2,..., (N - 1)$, with
mass $m_{i} = \mu_{2}$, form an (N - 2)-plet.
In that case, it is clear that the residual symmetry of the ground
state is (at least) O(N - 2), i.e., that the O(N) symmetry has been
dynamically (spontaneously) broken to (at least) O(N - 2). We say
``at least" because, when the masses $m_{0} = M = m_{1} = \mu_{1}$ are
equal there is additional O(2) symmetry of the vacuum.
Similar comments are valid in the case when more than two fields
develop v.e.v.s. Of course, the residual symmetry of the vacuum
determines the number of the NG bosons.
Next we turn to the case of two fields with v.e.v. as worked out in
Appendix \ref{s:ap1}.
We have shown in Appendix \ref{s:ap1} that there are only two distinct
mass solutions
$M = \mu_{1}, \neq \mu_{2}$ to Eqs. (\ref{e:gap2a}-\ref{e:gap2d}).
Thus the residual symmetry of the vacuum is $O(N-2) \times O(2)$, and
the corresponding number of NG bosons must be (2N - 4).
In the following we shall see how all these NG bosons come about.

\section{The Nambu-Goldstone theorem}
\label{s:NG}

\subsection{The two-body equation}
\label{s:bse}

In Ref. \cite{dms96,issei01} it was shown that in the Gaussian
approximation
to the O(2) and O(4) $\phi^4$ model, the Nambu-Goldstone particles
appear
in the {\it two-particle spectra}, i.e. that they are massless
bound states of two different massive elementary excitations
with an admixture of the (massive elementary) one-body state with
the same quantum numbers (the CDD pole). This admixture of the one-body
state is crucial for the masslessness of the NG state [it also proved
to be a source of confusion]. As
there are at most (N - 1) such elementary particles/CDD poles
\footnote{One field adopts a v.e.v. so it cannot create or destroy
single NG bosons.},
it appears that there can be at most (N - 1) NG bosons in this theory.
There are $N(N-1)/2$ (distinct) pairs of elementary particles in
this model, however,
and an equal number of distinct (potentially bound) two-body states.
Thus there can be at most $N(N-1)/2$ NG bosons, precisely the maximum
number allowed by the O(N) Lie group generator counting. Next we
must show that the number of ``broken symmetry generators" corresponds
precisely to the number of massless bound states and that the corresponding
N\" other currents are conserved.

The two-body equation of motion in the Gaussian approximation is
equivalent to the
four-point Schwinger-Dyson (or Bethe-Salpeter) equation [for proof
of this equivalence, see Ref. \cite{bg80,kerm98}].
All the NG channels obey a generic four-point SD equation that reads
\begin{eqnarray}
D_{ij,ij}(s) &=& V_{ij,ij}(s)
+ V_{ij,kl}(s) \Pi_{kl,mn}(s) D_{mn,ij}(s)
\label{e:bsepi} \
\end{eqnarray}
where $D(s)$ is the four-point Green function $(N \times N)$ matrix
(scattering amplitude), $\Pi(s)$ is the polarization
function matrix and $V(s)$ is the potential matrix;
$ij$ denote the O(N) indices of the two
constituents, with the generic solution
\begin{eqnarray}
{\bf D}(s) = {\bf V}(s) [1 - {\bf V}(s) \mbox{\boldmath$\Pi$}(s)]^{-1}
\label{e:sol}
\end{eqnarray}
where $s = (p_{1} + p_{2})^{2} \equiv P^{2}$ is the center-of-mass
(CM) energy. These matrices may reduce to a direct sum of submatrices
depending on the residual symmetry of the system.
Such effective two-body propagators can also be written in the
following form (see Ref. \cite{gross})
\begin{eqnarray}
D_{\alpha}(s) &\simeq&
{g_{\alpha}^{2} \over{s - m_{\alpha}^{2}}},
\label{e:proppi}
\end{eqnarray}
where $m_{\alpha}$ is the effective mass in channel $\alpha$.
The difference between various O(N) ``flavour" sectors appears in the
polarization functions $\Pi(s)$ and potentials $V(s)$: more specifically
in (N - 2) channels [$(i=0,j\subset 2,...N-1) \equiv M\mu_{2}$]
the said matrices are diagonal and have the form
\begin{eqnarray}
\Pi_{M\mu_{2}}(s)
&=& I_{M\mu_{2}}(s) =
i \hbar \int {d^{4} k \over (2 \pi)^{4}}
{1 \over {\left[k^{2} - M^{2} + i \epsilon \right]
\left[(k - P)^{2} - \mu_{2}^{2} + i \epsilon \right]}}
\label{e:pols2} \\
V_{M\mu_{2}}(s) &=& 2 \lambda_{0} \left[1 + \left({2 \lambda_{0}
\langle \phi_{0} \rangle^{2}
\over{s - \mu_{2}^{2}}}\right) \right]=
2 \lambda_{0} \left[1 + {M^{2} \over{s - \mu_{2}^{2}}}\right]
\label{e:pot1}\
\end{eqnarray}
and another (N - 2) channels
[$(i=1,j\subset 2,...N-1) \equiv \mu_{1}\mu_{2}$ have the form
\begin{eqnarray}
\Pi_{\mu_{1}\mu_{2}}(s)
&=& I_{\mu_{1}\mu_{2}}(s) =
i \hbar \int {d^{4} k \over (2 \pi)^{4}}
{1 \over {\left[k^{2} - \mu_{2}^{2} + i \epsilon \right]
\left[(k - P)^{2} - \mu_{1}^{2} + i \epsilon \right]}}
\label{e:pols1} \\
V_{\mu_{1}\mu_{2}}(s) &=& 2 \lambda_{0} \left[1 +
\left({2 \lambda_{0}
\langle \phi_{1} \rangle^{2}
\over{s - \mu_{2}^{2}}}\right) \right] =
2 \lambda_{0} \left[1 + {\mu_{1}^{2} \over{s - \mu_{2}^{2}}}\right]
\label{e:pot2} .\
\end{eqnarray}
The poles ${1 \over{s - \mu_{1,2}^{2}}}$ in Eqs. (\ref{e:pot1}),
(\ref{e:pot2}) are called the
Castillejo-Dalitz-Dyson [CDD] poles and correspond to one-particle
states in the theory.
Now due to  Eqs. (\ref{e:gap2a}-\ref{e:gap2d})
these two channels happen to be equivalent
in this case, but that degeneracy is accidental (for a different
example see Sect. \ref{s:th}).

\subsection{Massless (Nambu-Goldstone) two-body states}
\label{s:ng}

With an $O(N-2) \times O(2)$ residual symmetry there ought to be
(2N - 4) = (N - 2) + (N - 2) NG bosons.
This number of NG bosons quickly exceeds
the number of available fields in the Lagrangian: $2(N - 2) > (N - 1)$
for $N > 3$.
Thus, there aren't
enough scalar fields to provide for all the NG bosons in case
of nonstandard symmetry breaking. Does this mean that the NG theorem
breaks down in such a case?

As in Ref. \cite{dms96}, we shall prove
that at zero CM energy $P = 0$, the matrix
$[1 - V_{M\mu_{2}}(0) \Pi_{M\mu_{2}}(0)]$
vanishes. We use Eq. (\ref{e:gapb}) to write
\begin{eqnarray}
V_{M\mu_{2}}(0) &=& 2 {\lambda_{0}}
\left[1 - {M^{2} \over \mu_{2}^{2}} \right]
\nonumber \\
\Pi_{M\mu_{2}}(0)&=& \hbar \left({I_{0}(M) - I_{0}(\mu_{2})\over M^{2} -
\mu_{2}^{2}}\right)
\ ~ \label{e:polz2}
\end{eqnarray}
then use Eq. (\ref{e:gap1a}) to obtain the final result
\begin{eqnarray}
V_{M\mu_{2}}(0) \Pi_{M\mu_{2}}(0) &=& 1 \ ~.
\label{e:polz3}
\end{eqnarray}
Thus the inverse propagator Eq. (\ref{e:sol}) evaluated at zero
momentum vanishes:
\begin{eqnarray}
D_{M\mu_{2}}^{-1}(0) &=&
[1 - V_{M\mu_{2}}(0) \Pi_{M\mu_{2}}(0)] V_{M\mu_{2}}(0)^{-1} = 0,
\label{e:pimass}
\end{eqnarray}
which is equivalent to the result $m_{M\mu_{2}}^2 = 0$.
Similarly for the remaining (N - 2) $\mu_{1}\mu_{2}$ channels.
{\it q.e.d.}

Note that the proof in no way depended on the fact that $\mu_{1} = M$.
In other words, if the gap Eqs. (\ref{e:gap2a}-\ref{e:gap2d}) allowed
a solution such that $\mu_{1} \neq M$ and
$\langle \phi_{1} \rangle = 0$, the NG theorem would still hold.
But, then the CDD poles would decouple in the latter
(N - 2) channels, i.e., there would be no CDD poles in these channels.
This shows that, at least in principle, (N - 2) NG bosons could be
pure bound states with no single particle admixtures. In Sect.
\ref{s:th} we shall give an example of a model with a gap equation
that allows such solutions.

\subsection{Conservation of N\" other currents}
\label{s:axwti}

There are $a \subset 2, ..., N-1$ dynamically broken
O(N) symmetry N\" other current matrix elements corresponding to
\begin{eqnarray}
J_{\mu 5}^{a} (p^{'}, p) &=&
\langle \phi^{a} (p')|J_\mu(0)|\phi_{0}(p) \rangle
\nonumber \\
&=&
(p^{'} + p)_{\mu} + q_{\mu} \left({M^{2} \over q^{2} -
\mu_{2}^{2}}\right)
\nonumber \\
&-&
\Gamma_{\mu 5}^{M\mu_{2}} (q) D_{M\mu_{2}}(q) ~.
\label{e:ax1} \
\end{eqnarray}
where $\Gamma_{\mu 5}^{M\mu_{2}} (q)$ is
\begin{eqnarray}
\Gamma_{\mu 5}^{M\mu_{2}} (q) &=&
i \int {d^{4} k \over (2 \pi)^{4}}
\left[(2 k + q)_{\mu} + q_{\mu}
\left({M^{2} \over q^{2} - \mu_{2}^{2}}\right)
\right]
{1 \over {\left[k^{2} - M^{2} \right]
\left[(k + q)^{2} - \mu_{2}^{2} \right]}}
\nonumber \\
&=&
{q_{\mu} \over q^{2}}\left[{\mu_{2}^{2} \over{2 \lambda_{0}}}
\left(V_{M\mu_{2}}(0) \Pi_{M\mu_{2}}(0) - V_{M\mu_{2}}(q^{2})
\Pi_{M\mu_{2}}(q^{2}) \right)
\right] .\
\label{e:ax}
\end{eqnarray}
Inserting the vertex $\Gamma_{\mu 5}^{M\mu_{2}} (q)$, Eq. (\ref{e:ax})
together
with the two-body propagator $D_{M\mu_{2}}(q^{2})$, Eq. (\ref{e:sol})
into Eq. (\ref{e:ax1}) one finds
\begin{eqnarray}
J_{\mu 5}^{a} (p^{'}, p) &=&
(p^{'} + p)_{\mu} + q_{\mu}
\left({M^{2} \over q^{2} - \mu_{2}^{2}}\right)
\nonumber \\
&-&
{q_{\mu} \over q^{2}}
\left[{\mu_{2}^{2} \over{2 \lambda_{0}}}\right]
V_{M\mu_{2}}(q^{2})
\nonumber \\
&=& (p^{'} + p)_{\mu} + q_{\mu}
\left({M^{2} - \mu_{2}^{2} \over{q^{2}}}\right) ~,
\label{e:ax3} \
\end{eqnarray}
where
$q^{\nu} = (p^{'} - p)^{\nu}$.
This current is manifestly devoid of a pole at $q^{2} = \mu_{2}^2$.
The composite state plays precisely the role of the Nambu-Goldstone
boson in the conservation of the dynamically broken
O(N) symmetry N\" other currents \cite{njl61}, i.e.,
in the basic O(N) symmetry Ward-Takahashi identity, c.f. Refs.
\cite{lee69,sym70},
\begin{eqnarray}
q^{\nu} J_{\nu 5}^{a} (p^{'}, p)  &=&
\left(p^{'2} - \mu_{2}^{2}\right) -
\left(p^{2} - M^{2}\right) ~,
\label{e:elwi} \
\end{eqnarray}
that follows directly from Eq. (\ref{e:ax3}). Similarly for
the remaining N-2 [$a \subset 2, .... N-1$] dynamically broken
O(N) symmetry N\" other current matrix elements corresponding to
\begin{eqnarray}
J_{\mu 5}^{a} (p^{'}, p) &=&
\langle \phi^{a} (p')|J_\mu(0)|\phi_{1}(p) \rangle
\nonumber \\
&=&
(p^{'} + p)_{\mu} + q_{\mu} \left({\mu_{1}^{2} \over q^{2} -
\mu_{2}^{2}}\right)
\nonumber \\
&-&
\Gamma_{\mu 5}^{\mu_{1}\mu_{2}} (q) D_{\mu_{1}\mu_{2}}(q) ~.
\label{e:ax4} \
\end{eqnarray}
The identity of the two ``gap", or CDD masses $\mu_{1} = M$ and the
concomitant ``excess" O(2) vacuum
symmetry are consequences of the simplicity of the vacuum equation
(\ref{e:vac}) that only depends on one O(N) algebraic invariant
\cite{klein63,cicogna}.
That, in turn, is a consequence of
the fact that we are dealing with fields in the fundamental irrep.
of O(N) and the requirement that the Lagrangian (\ref{e:lag}) be
renormalizable, i.e. at most of the fourth power in the fields.
The assumption of a second v.e.v.
$\langle \phi_{3} \rangle = \langle \pi^{0} \rangle \neq 0$ is
particularly unrealistic in the Gell-Mann--Levy model (O(N=4)
$\phi^{4}$ model) \cite{gml60}, because
of the negative parity of the $\pi$ fields: their nonzero v.e.v.
would imply spontaneous breaking of P and CP ``parities".
In Sect. \ref{s:th} we give an example of a $\phi^4$ model with
an internal symmetry that leads to a gap equation
with distinct mass solutions and potentially exotic NG bound states.

\section{The 't Hooft model}
\label{s:th}

\subsection{Definition of the
model}
\label{s:defth}

't Hooft's \cite{th86} extension of the linear sigma model
Lagrangian reads
\begin{eqnarray}
{\cal L_{\rm tH}} &=&
{\rm tr} \left[ (\partial_{\mu} {\rm M} \partial^{\mu}
{\rm M}^{\dagger}) + \mu^{2} {\rm M} {\rm M}^{\dagger} \right]
\nonumber \\
&-&
{1 \over 2} \left(\lambda_{1} - \lambda_{2}\right)
\left[ {\rm tr} \left({\rm M} {\rm M}^{\dagger} \right)\right]^{2}
-
\lambda_{2}
{\rm tr} \left[ \left({\rm M} {\rm M}^{\dagger} \right)^{2}\right]
\nonumber \\ &+&
2 \kappa \left[e^{i \theta} {\rm det} {\rm M} + c.c. \right]~~~,\  ~
\label{e:th}
\end{eqnarray}
where
\begin{eqnarray}
{\rm M} &=& {1 \over \sqrt{2}} \left(\Sigma + i \Pi\right)
\nonumber \\
\Sigma &=& {1 \over \sqrt{2}} \left(\sigma +
\mbox{\boldmath$\alpha$} \cdot \mbox{\boldmath$\tau$}\right)
 \nonumber \\
\Pi &=& {1 \over \sqrt{2}} \left(\eta^{} + \mbox{\boldmath$\pi$}
\cdot \mbox{\boldmath$\tau$}\right)~~. \
\end{eqnarray}
Eq.(\ref{e:th}) is equivalent to the following
\begin{eqnarray}
 {\cal L_{\rm tH}} &=&
{1 \over 2} \big[ (\partial_{\mu} \sigma)^{2} +
(\partial_{\mu} \mbox{\boldmath$\pi$})^{2}
+ (\partial_{\mu} \eta)^{2} +
(\partial_{\mu} \mbox{\boldmath$\alpha$})^{2} \big]
\nonumber \\
&+&
 {\mu^{2} \over 2} \big[ \sigma^{2} + \mbox{\boldmath$\pi$}^{2}
+ \eta^{2} + \mbox{\boldmath$\alpha$}^{2}\big]
\nonumber \\
&+&
2 \kappa \cos\theta \big[\sigma^{2} + \mbox{\boldmath$\pi$}^{2}
- \eta^{2} - \mbox{\boldmath$\alpha$}^{2} \big]
\nonumber \\
&-&
4 \kappa \sin\theta \big[\sigma \eta - \mbox{\boldmath$\pi$}
\cdot \mbox{\boldmath$\alpha$} \big]
\nonumber \\
&-&
 {\lambda_{1} \over 8}\big[ \sigma^{2} + \mbox{\boldmath$\pi$}^{2}
+ \eta^{2} + \mbox{\boldmath$\alpha$}^{2} \big]^{2}
\nonumber \\
&-&
{\lambda_{2} \over 2}
\big[ \left(\sigma \mbox{\boldmath$\alpha$}
+ \eta^{} \mbox{\boldmath$\pi$} \right)^{2}
+ \left( \mbox{\boldmath$\pi$} \times \mbox{\boldmath$\alpha$} \right)^{2}
\big] \  ~
\label{bos}
\end{eqnarray}
which describes the dynamics of the two chiral meson quartets,
$(\sigma,\mbox{\boldmath$\pi$})$ and $(\mbox{\boldmath$\alpha$},\eta)$,
in this model, and
$\lambda_{1}, \lambda_{2}, \kappa, \theta$ are the bare
coupling constants. Nonvanishing angle $\theta$
leads to the explicit (not spontaneous) CP violation in
this model, so we set it equal to zero.
Thus we see that the 't Hooft model consists of two
coupled Gell-Mann--L\' evy (GML) linear sigma models \cite{gml60}, one
with a light and the other with a heavy quartet of mesons.

Note that the symmetries of various parts of the interaction Lagrangian
also vary, see App. \ref{s:symth}: (i)
\begin{eqnarray}
\lambda_{1} \neq 0;
\lambda_{2} = \kappa = \theta = 0
\label{e:coupl}
\end{eqnarray}
implies $O(8)$ symmetry.
(ii)
\begin{eqnarray}
\lambda_{1} \neq 0 \neq \lambda_{2}; \kappa = \theta = 0
\label{e:coup2}
\end{eqnarray}
implies $O(4) \times O(2)$ symmetry.
(iii)
\begin{eqnarray}
\lambda_{1} \neq 0 \neq \lambda_{2}; \kappa \neq 0 = \theta
\label{e:coup3}
\end{eqnarray}
implies $O(4)$ symmetry.
And the number of NG bosons must change accordingly.

\subsection{The gap equations}
\label{s:gapth}

The first set of energy minimization equations (\ref{e:vac}) reads
\begin{eqnarray}
0 &=& - v \left(\mu_{0}^{2} + 4 \kappa \right)
+ {\lambda_{1} \over 2} v \left[v^{2} +
3 I_{0}(m_{\sigma}) + 3 I_{0}(m_{\alpha})
+ 3 I_{0}(m_{\pi}) + I_{0}(m_{\eta})\right]
\nonumber \\
&+&
3 \lambda_{2} v I_{0}(m_{\alpha})
\label{e:vev1a} \\
m_{\sigma}^{2} &=&
- \mu_{0}^{2} - 4 \kappa +
{\lambda_{1} \over 2} v \left[v^{2} +
3 I_{0}(m_{\sigma}) + 3 I_{0}(m_{\alpha})
+ 3 I_{0}(m_{\pi}) + I_{0}(m_{\eta})\right]
\nonumber \\
&+&
3 \lambda_{2} I_{0}(m_{\alpha})
\label{e:gap1a} \\
m_{\alpha}^{2} &=&
- \mu_{0}^{2} - 4 \kappa + \left({\lambda_{1} \over 2}
+ \lambda_{2}\right) v^{2} +
{\lambda_{1} \over 2} \left[
I_{0}(m_{\sigma}) + 5 I_{0}(m_{\alpha})
+ 3 I_{0}(m_{\pi}) + I_{0}(m_{\eta})\right]
\nonumber \\
&+&
\lambda_{2} \left[I_{0}(m_{\sigma}) + 2 I_{0}(m_{\pi})\right]
\label{e:gap1b} \\
m_{\pi}^{2} &=&
- \mu_{0}^{2} - 4 \kappa + {\lambda_{1} \over 2} v^{2} +
{\lambda_{1} \over 2} \left[v^{2} +
I_{0}(m_{\sigma}) + 3 I_{0}(m_{\alpha})
+ 5 I_{0}(m_{\pi}) + I_{0}(m_{\eta})\right]
\nonumber \\
&+&
 \lambda_{2} \left[I_{0}(m_{\eta}) + 2 I_{0}(m_{\alpha})\right]
\label{e:gap1c} \\
m_{\eta}^{2} &=&
- \mu_{0}^{2} - 4 \kappa + {\lambda_{1} \over 2} v^{2} +
{\lambda_{1} \over 2} \left[v^{2} +
3 I_{0}(m_{\sigma}) + 3 I_{0}(m_{\alpha})
+ 3 I_{0}(m_{\pi}) + I_{0}(m_{\eta})\right]
\nonumber \\
&+&
3 \lambda_{2} I_{0}(m_{\pi})
\label{e:gap1d} \
\end{eqnarray}
where the divergent integral $I_{0}(m_{i})$ is given by Eq. (\ref{e:I_0}).
For simplicity in the following we use the following short-hand
notation $\alpha = m_{\alpha}^{2}; \eta = m_{\eta}^{2};
\sigma = m_{\sigma}^{2}; \pi = m_{\pi}^{2}.$
Eqs. (\ref{e:vev1a}-\ref{e:gap1d}) can be solved for
\begin{eqnarray}
\Pi_{\alpha \eta}(0) &=&
\frac{I_{0}(\alpha) - I_{0}(\eta)}{\alpha  - \eta} =
\frac{1}
{{\left({{{\lambda }_1}}^2 - {{{\lambda }_2}}^2 \right)}}
\left( {{\lambda }_1} + {{\lambda }_2} \left(\frac{\pi  - \sigma}
{\alpha  - \eta } \right) \right)
\nonumber \\
\Pi_{\pi \sigma}(0) &=&
\frac{I_{0}(\alpha) - I_{0}(\eta)}{\alpha  - \eta }=
\frac{- 1}
{{\left( {{{\lambda }_1}}^2 - {{{\lambda }_2}}^2 \right)}}
\left({{\lambda }_1} \, \frac{\pi}
{\sigma - \pi} + {{\lambda }_2}\, \frac{
\left( \alpha  - \eta  - \frac{{{\lambda }_2}}
{{{\lambda }_1}}  \, \sigma\right) }{\sigma - \pi} \right)
\nonumber \\
\Pi_{\pi \alpha}(0) &=&
\frac{I_{0}(\alpha) - I_{0}(\pi)}{\alpha  - \pi}
\nonumber \\
&=&
\frac{1}
{{\left({{\lambda }_1}^2 - 2 {{\lambda }_1}{{\lambda }_2}
- 3 {{{\lambda }_2}}^2 \right)}}
\left({{\lambda }_1} \, \left(1 - \frac{8 \kappa}
{\alpha - \pi}\right) + {{\lambda }_2}\, \frac{
\eta  - 8 \kappa}{\alpha - \pi} \right)
\nonumber \\
\Pi_{\sigma \eta}(0) &=&
\frac{I_{0}(\sigma) - I_{0}(\eta)}{\sigma  - \eta }
\nonumber \\
&=&
\frac{1}
{{\left({{\lambda }_1}^2 - 2 {{\lambda }_1}{{\lambda }_2}
- 3 {{{\lambda }_2}}^2 \right)}}
\left(\left({{\lambda }_1} - 2 {{\lambda }_2} \right)\,
\left(\frac{8 \kappa - \eta}
{\sigma - \eta}\right) - 3 {{\lambda }_2}\, \frac{
\alpha  - \pi - 8 \kappa - \frac{{{\lambda }_2}}
{{{\lambda }_1}}  \, \sigma }{\sigma - \eta} \right)
\ ~ \label{e:polz1}
\end{eqnarray}
whence follows
\begin{eqnarray}
\Pi_{\pi \eta}(0) &=&
\frac{I_{0}(\pi) - I_{0}(\eta)}{\pi  - \eta }
\nonumber \\
&=&
\frac{1}
{{\lambda }_1 \left(\pi - \eta \right)} \left(
8 \kappa + \pi
- \eta + {\lambda }_2 \left({\alpha - \eta}\right)
\Pi_{\eta \alpha}(0) - 3 {\lambda }_2 \left({\alpha - \pi}\right)
\Pi_{\pi \alpha}(0) \right)
\nonumber \\
\Pi_{\alpha \sigma}(0)
&=&
\frac{I_{0}(\sigma) - I_{0}(\alpha)}{\sigma  - \alpha }
\nonumber \\
&=&
\frac{1}
{{\lambda }_1 \left(\sigma - \alpha \right)} \left(
8 \kappa + \frac{{{\lambda }_2}}{{{\lambda }_1}}  \, \sigma
- \alpha + {\lambda }_2 \left({\sigma - \pi}\right)
\Pi_{\pi \sigma}(0) - 3 {\lambda }_2 \left({\alpha - \pi}\right)
\Pi_{\pi \alpha}(0) \right)
\ ~ \label{e:polz4}
\end{eqnarray}

\subsection{The Nambu-Goldstone theorem in the isotensor
pseudoscalar sector}
\label{s:ngth}

Note that there are $8 \times 8 = 64$ possible initial or final
two-body states here.
Thus there are $64 \times 64 = 4096$ possible channels, but
only $4 \times 7 = 28$ distinct pairs of particles,
or equivalently at most 28 possible NG bosons.
The last statement holds under the proviso of exact $O(8)$ symmetry
being broken to a discrete (non Lie) symmetry, however. Otherwise there
are fewer than 28 NG bosons, and when O(8) is explicitly broken down
to $O(4) \times O(2)$, or $O(4)$ by one of the terms in the Lagrangian
Eq. (\ref{e:th}), there are even fewer than that. This fact tells us
that many channels must be coupled and that the residual symmetry plays
a crucial role in this coupling.
We shan't look at every possible channel in this paper, but rather
concentrate only on the exotic ones, in this case the isotensor, so
as to show the existence of pure bound state NG bosons without CDD
admixtures.


We may use the residual vacuum symmetry, e.g. the $O(3)$ isospin
invariance to split this
$64 \times 64$ matrix equation into six invariant subspaces: three
flavour channels [(a) isoscalar; (b) isovector, and (c) isotensor]
of either parity.
In the two-body, or Bethe-Salpeter (BS) equation for the four-point
Green functions $D_{ij}(s)$, the indices $i,j$ denote the isospin of
the two-body
initial and final states, respectively.


The negative parity (pseudoscalar) isotensor two-body equation
is a single-channel one and straightforward to solve, see
Eq. (\ref{e:sol}). We
look at the zero CM energy $P = 0$ function $V_{\pi\alpha}(0)
\Pi_{\pi \alpha}(0) $; we use
\begin{eqnarray}
V_{\pi \alpha}^{(I=2)} &=& {{\lambda }_1} + {{\lambda }_2}
\ ~ \label{e:pot3}
\end{eqnarray}
and Eq. (\ref{e:polz1}) to obtain the final result
\begin{eqnarray}
V_{\pi \alpha}^{(I=2)}(0) \Pi_{\pi\alpha}(0) &=&
- \left(\frac{{{\lambda }_1} + {{\lambda }_2}}
{{{{\lambda }_1}^2 - 2 {{\lambda }_1}{{\lambda }_2}
- 3 {{{\lambda }_2}}^2 }}\right)
\left({{\lambda }_1} \, \left(1 - \frac{8 \kappa}
{\alpha - \pi}\right) + {{\lambda }_2}\, \frac{\eta  - 8 \kappa}
{\alpha - \pi} \right)
\ ~.
\label{e:polz5}
\end{eqnarray}
Now set $\lambda_{2} = \kappa \to 0$ and find
\begin{eqnarray}
\lim_{\lambda_{2} = \kappa \to 0}
V_{\pi \alpha}^{(I=2)}(0) \Pi_{\pi\alpha}(0) &=& 1 \ ~.
\label{e:polz6}
\end{eqnarray}
The propagator Eq. (\ref{e:sol}) evaluated at zero momentum
can be written as
\begin{eqnarray}
D_{\pi \alpha}^{(I=2)}(0) &=&
{g_{\pi \alpha}^{2} \over{- m_{\pi \alpha}^{2}}}
= {V_{\pi \alpha}(0) \over{1 - V_{\pi \alpha}(0)
\Pi_{\pi \alpha}(0)}} ,\
\label{e:istmass}
\end{eqnarray}
whence it follows that
\begin{eqnarray}
m_{\pi \alpha}^{2} &=&
\left[\frac{1}{{{{\lambda }_1} + {{\lambda }_2}}}
- \left(\frac{1}
{{{{\lambda }_1}^2 - 2 {{\lambda }_1}{{\lambda }_2}
- 3 {{{\lambda }_2}}^2 }}\right)
\left({{\lambda }_1} \, \left(1 - \frac{8 \kappa}
{\alpha - \pi}\right) + {{\lambda }_2}\, \frac{\eta  - 8 \kappa}
{\alpha - \pi} \right)\right]
\nonumber \\
&\times&
\left(\frac{\alpha - \pi}
{v}\right)^{2}
\nonumber \\
&=&
{\cal O}({\lambda }_2) + {\cal O}(\kappa)
\label{e:istm2}
\end{eqnarray}
Thus we see that the effective (pseudo) NG boson mass in this channel
is proportional to $\lambda_{2}$, and/or $\kappa$, the two O(8)
symmetry breaking parameters. {\it q.e.d.}

\section{Summary and Conclusions}
\label{s:s&c}

In summary, we have:
1) proven the NG theorem in the variational Gaussian wave functional
approximation to the O(N) symmetric $\phi^{4}$ model when the symmetry
of the ground state is a proper subgroup of O(N-1);
2) proven conservation of N\" other currents corresponding to the
dynamically broken symmetries;
3) proven the same NG theorem in the exotic isotensor channel of the
't Hooft model in the limit $\lambda_{2} = \kappa \to 0$.
The NG bosons are massless bound states of two massive constituents.
We emphasize that our proofs do not depend on the
specific values of the bare parameters, or of the cutoff in the
theory, so long as the system is in the spontaneously broken phase
with appropriate symmetry.

We should like to put these results into their proper logical and
chronological setting.
The variational method in quantum field theory (QFT) is based
on the Schr\" odinger representation and goes by the name of
Gaussian approximation to the ground state wave functional.
This method, in its various guises, was pioneered by Schiff,
Rosen and Kuti \cite{schiff63,rosen68,cjt74}in the 1960's and 70's,
and later revived and elaborated in the 1980's by Barnes and Ghandour
\cite{bg80,hat92}, and by Symanzik \cite{sym81} and by Consoli,
Stevenson and collaborators \cite{stev84,stev87,bc85}.
Related formalisms based on effective potentials and other functional
methods were discussed in Ref. \cite{cjt74,bc85,stev87} and references
cited therein.
Most of these studies addressed the $\phi^{4}$ scalar field theory
that is also the prime example of the Nambu-Goldstone (NG) theorem
\cite{njl61,gold61,gsw62,klein63,gilbert64},
an exact result in the O(N) symmetric $\phi^{4}$ scalar theory with
spontaneous internal symmetry breaking.

The NG theorem
was first shown {\it not} to be satisfied by the solutions to the
mass, or ``gap" equations in the Gaussian approximation by
Kamefuchi and Umezawa in 1964 \cite{ku64}, practically simultaneously
with the general proofs of the NG theorem
\cite{gsw62,klein63,gilbert64}.
This fact was subsequently rediscovered several
times \cite{hung66,bc85} and this unsatisfactory situation
persisted until 1994 \cite{dms96}. Various conjectures
as to the reasons for this failure and as to potential remedies were
advanced during this period of time.
It was first shown in Ref. \cite{dms96} that this apparent breakdown
of the NG theorem is not an intrinsic shortcoming of
the Gaussian wave functional approximation, but rather a
consequence of incomplete previous analyses. In other words,
the NG theorem is satisfied, but the NG bosons are not excitations
of the elementary scalar fields, as initially expected.
Rather, they are massless bound states of two massive elementary
scalar excitations, in close analogy with Nambu's \cite{njl61}
proof of the NG theorem in a self-interacting fermion theory.
NG bosons are solutions to the two-body (or Bethe-Salpeter [BS])
equation in the Gaussian approximation. That equation was only rarely
considered in the literature \cite{rosen68,bg80}, and never before
Ref. \cite{dms96} in
the context of spontaneous symmetry breaking of purely bosonic models.

In this light the result is simple enough to understand, yet it
drew strong, albeit unpublished criticism and affirmation \cite{vd94}.
Perhaps the underlying reason for the misunderstanding by some was the
implication of the proof that the $\phi^4$ scalar field theory
could have bound states, which, as ``everybody knew" \cite{vd94},
disagrees with various ``rigorous no-go" and ``triviality" theorems
in the same theory \cite{book93}
\footnote{For more recent results comparing constructive QFT
to the Gaussian approximation in 1+1 dimensional field theories,
see Ref. \cite{Prodan}}.
The said theorems hold only in the limit of an infinite
cutoff, however, in which the Gaussian approximation also becomes
trivial \cite{stev84}.
For finite cutoffs, on the other hand, this is a
nontrivial theory that may contain bound states.

Soon after the first proof in Ref. \cite{dms96} it was also shown
along the lines of Ref. \cite{gsw62} that the NG theorem also follows
from the Gaussian effective potential, Ref. \cite{okop96},
but that proof did
not shed much light on the mechanisms that made the NG bosons come
about. Only later, in Ref. \cite{vd98,issei02}, it was explicitly
shown how this
formal proof relates to the Gaussian two-body equations of motion.
Another source of confusion was the apparent doubling of degrees of
freedom, at least in some ``flavour" (internal symmetry) channels,
{\it viz.} the existence
of massive ``elementary" and massless bound states in the same channel.
This problem was resolved in Ref. \cite{vd98}, wherein the
K\" allen-Lehmann spectral
function was calculated in appropriate channels of the model.
This spectral function clearly shows the
presence of massless NG states and the absence of
the massive single-particle excitations.
That also constitutes a proof of the NG theorem along Gilbert's lines, Ref.
\cite{gilbert64}, within the Gaussian wave functional approximation.
Thus we have confirmed all the well known proofs of the NG
theorem in the Gaussian approximation.

The NG theorem is the simplest example of a Ward-Takahashi identity,
which follows from the underlying internal
symmetry of the $\phi^4$ model. Ward-Takahashi identities
typically relate (n-1)-point
Green functions to n-point functions and/or matrix elements
of N\" other currents. These identities were developed by Lee in
the linear sigma model
at the perturbative one loop level \cite{lee69}, and by Symanzik for
arbitrary orders of perturbation theory \cite{sym70}, so
we shall call them the Lee-Symanzik [LS] identities.

The {\it exact}, i.e. nonperturbative Green functions satisfy an
infinite set of coupled integro-differential equations called the
Schwinger-Dyson equations \cite{chai79}. The iterative/perturbative
solutions to the SD equations form (infinite/finite) sets of Feynman
diagrams.
If one decouples the SD equations for higher order Green functions
from the lower order ones
(in popular jargon, if one truncates the SD equations), one may obtain
tractable equations and find their solutions that sum infinite, albeit
incomplete sets of Feynman diagrams. It has been known at least since
1980, Ref. \cite{bg80}, that the Gaussian approximation to the unbroken
symmetry $\phi^4$ theory corresponds to one such truncation of the SD
equations. But, truncated SD equations
need not obey the conservation laws of the original
SD equations of motion, i.e. LS identities may be violated by the
truncation. To our
knowledge, no proof of LS identities had been given for truncated
SD equations, i.e. for infinite
classes of diagrams in the bosonic linear sigma model prior to Ref.
\cite{dms96}, although a similar proof had been given by Nambu
and Jona-Lasinio in their fermionic model some 30 years before
\cite{njl61}.
Thus we have shown that the Gaussian functional approximation constitutes
a closed, self-consistent symmetry-preserving approximation to the
Schwinger-Dyson equations.


\section*{Acknowledgements}

One of the authors [V.D.] would like to acknowledge a center-of-excellence
(COE) Professorship for the year 2000/1 and the hospitality of RCNP.
The same author also wishes to thank Prof. Paul Stevenson for
valuable conversations and correspondence relating to the Gaussian
functional approximation.

\appendix{}
\section{The gap equations with two v.e.v.s}
\label{s:ap1}

The first set of
energy minimization equations (\ref{e:vac}) read
\begin{eqnarray}
\mu_{0}^{2} &=&
{\lambda_{0}} \left[v^{2} +
3 \hbar I_{0}(M) + \hbar I_{0}(\mu_{1})
+ (N-2) \hbar I_{0}(\mu_{2}) \right]
\label{e:veva} \\
\mu_{0}^{2} &=&
{\lambda_{0}} \left[v^{2} + \hbar I_{0}(M)
+ 3 \hbar I_{0}(\mu_{1})
+ (N-2) \hbar I_{0}(\mu_{2}) \right]
\label{e:vevb} \\
v^{2} &=& \langle \phi_{0} \rangle^{2} +
\langle \phi_{1} \rangle^{2} = \langle \mbox{\boldmath$\phi$}
\rangle^{2} \\
\langle \phi_{i} \rangle &=& 0 ~~~i = 2, \ldots ,N-1~\ ~,
\label{e:vevc}
\end{eqnarray}
where the divergent integral $I_{0}(m_{i})$, Eq. (\ref{e:I_0}) is
understood to be
regularized via an UV momentum cut-off $\Lambda$, either three-, or
four dimensional.
The second set of gap equations, (\ref{e:vac}) read:
\begin{eqnarray}
M^{2} &=&
- \mu_{0}^{2} + {\lambda_{0}}
\left[2 \langle \phi_{0} \rangle^{2}
+ \langle \mbox{\boldmath$\phi$} \rangle^{2}
+ 3 \hbar I_{0}(M)  + \hbar I_{0}(\mu_{1}) +
(N - 2) \hbar I_{0} (\mu_{2}) \right]
\label{e:gapa} \\
\mu_{1}^{2} &=&
- \mu_{0}^{2} + {\lambda_{0}}
\left[2 \langle \phi_{1} \rangle^{2}
+ \langle \mbox{\boldmath$\phi$} \rangle^{2}
+ \hbar I_{0}(M)  + 3 \hbar I_{0}(\mu_{1}) +
(N - 2) \hbar I_{0} (\mu_{2}) \right]
\label{e:gapb} \\
\mu_{2}^{2} &=&
- \mu_{0}^{2} +
{\lambda_{0}} \left[\langle \mbox{\boldmath$\phi$} \rangle^{2}
+ \hbar I_{0}(M)  + \hbar I_{0}(\mu_{1}) +
N \hbar I_{0}(\mu_{2}) \right]
\label{e:gapc} \ ~.
\end{eqnarray}
Upon inserting Eqs. (\ref{e:veva},b) into Eqs. (\ref{e:gapa},b),
the following coupled ``gap" equations emerge:
\begin{eqnarray}
M^{2} &=&
2 {\lambda_{0}} \langle \phi_{0} \rangle^{2} + 2 \lambda_{0}
\hbar \left[I_{0}(M) - I_{0}(\mu_{1})\right] =
2 {\lambda_{0}} \langle \phi_{0} \rangle^{2}
\label{e:gap3a} \\
\mu_{1}^{2} &=&
2 {\lambda_{0}} \langle \phi_{1} \rangle^{2} - 2 \lambda_{0}
\hbar \left[I_{0}(M) - I_{0}(\mu_{1})\right] =
2 {\lambda_{0}} \langle \phi_{1} \rangle^{2}
\label{e:gap3b} \\
M^{2} - \mu_{1}^{2} &=&
2 {\lambda_{0}} \left(\langle \phi_{0} \rangle^{2}
- \langle \phi_{1} \rangle^{2}
+ 2 \lambda_{0}
+ \hbar \left[I_{0}(M) - I_{0}(\mu_{1})\right]\right)
\label{e:gap3c} \\
\mu_{2}^{2} &=&
2 \lambda_{0} \hbar \left[I_{0}(\mu_{2}) - I_{0}(M) \right] =
2 \lambda_{0} \hbar \left[I_{0}(\mu_{2}) - I_{0}(\mu_{1}) \right] ~. \
\label{e:gap3d}
\end{eqnarray}
Note that these equations lead to
\begin{eqnarray}
I_{0}(M) - I_{0}(\mu_{1}) &=& 0
\label{e:gap2a} \\
M^{2} - \mu_{1}^{2} &=&
2 {\lambda_{0}} \left(\langle \phi_{0} \rangle^{2}
- \langle \phi_{1} \rangle^{2}\right) = 0
\label{e:gap2c} ~,\
\end{eqnarray}
which, in turn have at least one solution:
$\mu_{1} = M$ and
$\langle \phi_{0} \rangle^{2} = \langle \phi_{1} \rangle^{2}$,
that are solutions to a single nontrivial gap equation
\begin{eqnarray}
M^{2} &=&
2 {\lambda_{0}} \langle \phi_{1} \rangle^{2}
\label{e:gap2b} \\
\mu_{2}^{2} &=&
2 \lambda_{0} \hbar \left[I_{0}(\mu_{2}) - I_{0}(M) \right]
~, \
\label{e:gap2d}
\end{eqnarray}
with two unknowns one of which is kept fixed, c.f. Ref. \cite{issei01}.
This gap equation has been solved numerically in Ref. \cite{issei01}:
it admits only massive solutions $M > \mu > 0$, however,
for real, positive values of $\lambda_{0}, \mu_{0}^{2}$ and real
ultraviolet cut-off $\Lambda$ in the momentum integrals $I_{0}(m_{i})$.
In other words the ``would be NG boson fields" ($\phi_{1,2,...,N-1}$)
excitations are all massive ($\mu > 0$) in the MFA.
This looks like a breakdown of the NG theorem in this approximation,
but, as discussed in Ref. \cite{dms96,issei01},
there is a solution by way of the two-body (Bethe-Salpeter) equation.

\section{Symmetries of the 't Hooft model}
\label{s:symth}

The two field quartets, $(\sigma,\mbox{\boldmath$\pi$})$ and
$(\mbox{\boldmath$\alpha$},\eta)$, have different ``chiral"
$O(4) = O(3) \times O(3) \simeq SU_{L}(2) \times SU_{R}(2)$
\begin{eqnarray}
\delta_{5} \sigma &=& \mbox{\boldmath$\beta$} \cdot
\mbox{\boldmath$\pi$} \\
\delta_{5} \eta &=& - \mbox{\boldmath$\beta$} \cdot
\mbox{\boldmath$\alpha$}
\\
\delta_{5} \mbox{\boldmath$\alpha$} &=&
\mbox{\boldmath$\beta$} \eta
\\
\delta_{5} \mbox{\boldmath$\pi$} &=& -
\mbox{\boldmath$\beta$} \sigma
~~, \
\label{e:chitrf}
\end{eqnarray}
isospin
\begin{eqnarray}
\delta \sigma &=& 0
\\
\delta \eta &=& 0
\\
\delta \mbox{\boldmath$\alpha$} &=& -
\mbox{\boldmath$\varepsilon$} \times \mbox{\boldmath$\alpha$}
\\
\delta \mbox{\boldmath$\pi$} &=& -
\mbox{\boldmath$\varepsilon$} \times \mbox{\boldmath$\pi$}
~~, \
\label{e:isotrf}
\end{eqnarray}
and $U_{A}(1) \simeq O(2)$ transformation properties
\begin{eqnarray}
\delta_{5}^{0} \sigma &=& \beta \eta
\\
\delta_{5}^{0} \eta &=& - \beta \sigma
\\
\delta_{5}^{0} \mbox{\boldmath$\alpha$} &=&
\beta \mbox{\boldmath$\pi$}
\\
\delta_{5}^{0} \mbox{\boldmath$\pi$} &=& -
\beta \mbox{\boldmath$\alpha$}
~~. \
\label{e:u(1)trf}
\end{eqnarray}
The Lie algebra O(4) has two Casimir operators, but there is only
one invariant in the $({1 \over 2},{1 \over 2})$ representation with
one meson quartet, {\it viz.} (i)
$\big[\sigma^{2} + \mbox{\boldmath$\pi$}^{2} \big]$, whereas
with two quartets one has three invariants: (i) above; (ii)
$\big[ \eta^{2} + \mbox{\boldmath$\alpha$}^{2} \big]$, and (iii)
$\big[ \eta \sigma - \mbox{\boldmath$\alpha$} \cdot
\mbox{\boldmath$\pi$} \big]$. Any odd power of the third invariant
(iii) violates CP. Even without the third invariant the algebraic
structure of the Lagrangian
is rich enough to allow for multiple vacua [solutions to the
energy minimization equations (\ref{e:min1})]
even in the (first) Born approximation.
For example, equations (\ref{e:min1}) applied directly to the 'tHooft
interaction (\ref{e:th}) in the Born approximation allow two nonzero
v.e.v.s
[$\langle \sigma \rangle_{0 {\rm B}} = v_{0 {\rm B}} \neq 0$ and
$\langle {\alpha }_{3} \rangle_{0 {\rm B}} = v_{1 {\rm B}} \neq 0$]:
\begin{eqnarray}
- \mu_{0}^{2} - 4 \kappa
+ {\lambda_{1} \over 2} \left[v_{0 \rm B}^{2} +
v_{1 \rm B}^{2} \right] + \lambda_{2} v_{1 \rm B}^{2} &=& 0
\label{e:Bvev1a} \\
- \mu_{0}^{2} + 4 \kappa
+ {\lambda_{1} \over 2} \left[v_{0 \rm B}^{2} +
v_{1 \rm B}^{2} \right] + \lambda_{2} v_{1 \rm B}^{2}
&=& 0 .
\label{e:Bvev1b}\
\end{eqnarray}
Their solutions are
\begin{eqnarray}
\langle \sigma \rangle_{0 {\rm B}} = v_{0 {\rm B}} =
{\mu_{0}^{2} \over{\lambda_{1} + \lambda_{2}}}
- 4 {\kappa \over{\lambda_{2}}}
\label{e:Bvev2a} \\
\langle {\alpha }_{3} \rangle_{0 {\rm B}} = v_{1 {\rm B}} =
{\mu_{0}^{2} \over{\lambda_{1} + \lambda_{2}}}
+ 4 {\kappa \over{\lambda_{2}}}
\label{e:Bvev2b}\
\end{eqnarray}
leading to a nontrivially broken ground state.
Note, however, that (i) $\lambda_{2} \to 0$ is a singular limit point,
i.e. these vacua are not continuously connected with the more
conventional vacua at $\lambda_{2} = 0$, and
(ii) the two kinds of vacua/energy minima coincide whenever
$\kappa = 0$ and ${\lambda}_{2} \neq 0$.
Hence it is possible for the system to be in an unconventional vacuum
even with infinitesimally small
${\lambda}_{2}$ and $\kappa$ when they vanish with a fixed nonzero
ratio. This makes it plausible that unusual vacua may also appear
in this model when the symmetry is broken dynamically, i.e. by
``loop" effects. This doubling of vacua is a consequence of
multiple (two) independent algebraic invariants in the Lagrangian
(\ref{e:th}) which in turn led to multiple (two) v.e.v.s, in
agreement with the general theory \cite{klein63,cicogna}.

\end{document}